\documentclass[%
reprint,
amsmath,
amssymb,
aps,
pra,
superscriptaddress,
preprintnumbers,
prstab,
prstper,groupedaddress]{revtex4-2}
\usepackage{graphicx}
\usepackage{dcolumn}
\usepackage{bm}

\usepackage[english]{babel}
\usepackage[utf8]{inputenc}
\usepackage{amssymb,amsfonts,amsmath,mathtools,mathrsfs}
\usepackage{relsize}
\usepackage{mdframed}
\usepackage{enumitem}
\usepackage{graphicx}
\usepackage{url,hyperref}
\usepackage[table,usenames,dvipsnames]{xcolor}
\usepackage{gensymb}
\usepackage{grffile}
\usepackage{float}

\hypersetup{
colorlinks=true,
linkcolor = blue,
filecolor = blue,
urlcolor = Blue,
citecolor = blue,
pdftitle = {ssSSB_Arxiv}
}

\usepackage[capitalize]{cleveref}
\usepackage{textcomp}
\usepackage{braket} 

\begin{document}

\title{Site Selective Spontaneous Symmetry Breaking and Partial Order in Kondo Lattices}

\author{Soumyaranjan Dash}
\author{Sanjeev Kumar}
\affiliation{Department of Physical Sciences, Indian Institute of Science Education and Research (IISER) Mohali, Sector 81, S.A.S. Nagar, Manauli PO 140306, India}

\begin{abstract}
Using the combination of a new effective Hamiltonian approach and hybrid Monte-Carlo simulations, we unveil a variety of partially magnetically ordered (PMO) phases in the Kondo lattice model. Our approximation is motivated by two crucial features of the Hamiltonian: (i) formation of Kondo singlets leading to vanishing local magnetic moments, and (ii) spatially correlated nature of the effective single-particle kinetic energy. We discover PMO phases with fractional values $1/4$, $3/8$, and $1/2$ of Kondo-screened sites. A common understanding of these states emerges in terms of a non-local ordering mechanism. The concept of site-selective spontaneous symmetry breaking introduced here provides a new general approach to study models of interacting fermions in the intermediate coupling regime.
\end{abstract}

\date{\today}
\maketitle

{\it Introduction:--}
The Kondo impurity model (KIM) and the Kondo lattice model (KLM) are two of the cornerstones of strong correlation physics \cite{Kondo1964, Hewson1993, Coleman2015}. These models are commonly employed to describe the physics of conduction electrons in the presence of magnetic impurities. While the KIM was introduced to understand the resistivity minimum in metals with magnetic impurities \cite{Kondo1964, DeHaas1934, Sarachik1964, Kondo2013}, the KLM emerged as its natural extension for modelling materials that host a lattice of localized spins, such as the heavy fermion compounds \cite{Hewson1993, Stewart1984a, Wirth2016}. In addition to the electrical transport properties, it is important to understand the nature of magnetic order in such models and materials. In fact, in some cases the magnetic order is known to non-trivially influence the electrical transport \cite{Nagaosa2010, Martin2008, Kumar2010, Ishizuka2013b, Taguchi2009, Venderbos2012}.
In the limit of small Kondo coupling, the magnetic order is described via the well-known Ruderman-Kittel-Kasuya-Yosida (RKKY) interactions mediated via the conduction electrons \cite{Ruderman1954, Kasuya1956, Yosida1957}. On the other hand, a natural starting point to understand the large Kondo-coupling limit is to invoke the concept of singlet formation between conduction electrons and local spins. In general, the competition between the kinetic energy of electrons and that of the singlet formation dictates the balance between the RKKY ordered magnets and non-magnetic metallic or insulating states having all localized spins participate in singlet formation \cite{Pruser2014, Spinelli2015}. This is summarized in the Doniach phase diagram for heavy fermion compounds \cite{Doniach1977}.

Recent experiments report an intriguing intermediate possibility. The extent to which a magnetic moment is screened by conduction electrons can be site-dependent \cite{Iyer2023, Lucas2017, Moro-Lagares2019, Giannakis2019, Pfleiderer2004, Movshovich1999}. This leads to a remarkable set of new magnetically ordered phases where Kondo singlets and magnetic moments not only co-exist but also spatially order. This possibility is beyond the Doniach phase diagram of Kondo systems, and has become an active topic of theoretical investigations in recent years \cite{Costa2017, Aulbach2015, Kruger2012, Ishizuka2012b, Ishitobi2023, Bernhard2015, Kim2022, Kavai2021, Fahl2021, Kessler2020, Peters2017, Peters2012, Motome2010a, Sato2018}. Most of the efforts to quantitatively understand such novel magnetic order have relied on the use of computationally demanding methods, such as, the cluster dynamical mean-field theory (cDMFT) or quantum Monte Carlo (QMC) \cite{Peters2017, Peters2012, Motome2010a, Sato2018}. While the results from these methods provide good benchmark for any approximations, often the underlying mechanisms remain unclear.

In this work, we present a new approach to discover partially magnetically ordered (PMO) phases of the KLM. The approximation combines the key features of the KLM in the weak and strong coupling limits. The concept of site-selective spontaneous symmetry breaking (ssSSB) is introduced to allow for non-trivial spatial correlations and partial ordering. Important parameters of the effective Hamiltonian are explicitly computed via exact diagonalization (ED). We implement the hybrid Monte Carlo (HMC) method to investigate the model at half and quarter filling. The HMC results supplemented by variational calculations show that four novel PMO phases, characterized via spin and singlet structure factors, exist as the ground states at intermediate Kondo coupling strengths. We present finite temperature phase diagrams of the model and highlight some common features between the phases obtained in our work and those reported in the experiments on heavy fermion compounds. Our results qualitatively extend the Doniach phase diagram to include new phases in the intermediate Kondo coupling and intermediate temperature regimes. 

{\it Motivating the effective Hamiltonian:--}
Consider two spin-1/2 impurities Kondo-coupled to a conduction band \cite{Trishin2023, Jones1987, Jones1988, Jones1989, Spinelli2015, Jayaprakash1981}. The corresponding Hamiltonian in the tight-binding formulation is given by,
\begin{eqnarray}
	H & = & - t \sum_{\langle ij \rangle,\alpha} (c^\dagger_{i\alpha} c^{}_{j\alpha} + {\textrm H.c.})
	+ J_{\text{K}} \sum_{i \in \{\ell ,m\}} {\bf S}_i \cdot {\bf s}_i, 
	\label{eq:Ham1}
\end{eqnarray}
\noindent
where, $c^{}_{i\alpha} (c_{i\alpha}^\dagger)$ annihilates (creates) an electron at site ${i}$ with spin 
$\alpha$. The first term represents electronic kinetic energy quantified by the nearest neighbor (nn) hopping $t$, which also sets the basic energy scale. The second term describes Kondo coupling between localized quantum spins ${\bf S}_{\ell}$ and ${\bf S}_m$ of magnitude $1/2$ and the electronic spin ${\bf s}_i = \frac{1}{2} \sum_{\alpha, \beta} c_{i\alpha}^\dagger \boldsymbol{\sigma}_{\alpha \beta} ~c^{}_{i\beta}$, where $\boldsymbol{\sigma}$ is Pauli matrix vector and $J_{\rm K}$ is the coupling constant. 
At sufficiently low temperatures, isolated impurities are screened by the conduction electrons and the magnetic moments vanish. The RKKY interactions dominate at small $J_{\rm K}$ and, in the case of the KLM, a macroscopic magnetic order parameter develops through the SSB mechanism and the low-temperature quantum effects are masked. However, for strong coupling, the Kondo scale dominates and the magnetic order is suppressed. This homogeneous picture clearly excludes the possibility of partial order reported in many experiments \cite{Iyer2023, Lucas2017, Moro-Lagares2019, Giannakis2019, Pfleiderer2004, Movshovich1999}. We propose the following effective single-particle Hamiltonian to allow for inhomogeneous phases with partial order:

\begin{eqnarray}
	H & = & - t \sum_{\langle ij \rangle,\sigma} (c^\dagger_{i\sigma} c^{}_{j\sigma} + {\textrm H.c.})
	+ J_{\text{K}} \sum_{i} (1-\lambda_i){\mathbf{\cal S}}_i \cdot {\bf s}_i \nonumber \\
	& & + J_{\text{K}} \sum_{i} \lambda_i c^\dagger_{i\sigma} c^{}_{i\sigma} + E_{\text{K}} \sum_{i} \lambda_i + \sum_{ij} V_{ij} \lambda_i \lambda_j. 
	\label{eq:Ham2}
\end{eqnarray}

\noindent
In the above, we have introduced the idea of ssSSB by allowing a localized spin to assume a dual character: it captures a conduction electron to form a Kondo singlet with probability $\lambda_i$, and appears as a classical spin with probability $(1-\lambda_i)$ \cite{Nakatsuji2004}. Accordingly, $E_{\rm K}  = \frac{-3J_{\rm K}}{4}\big( \dfrac{1-e^{-J_{\rm K}/T}}{1+3e^{-J_{\rm K}/T}} \big)$ and $V_{ij}$ in Eq. (\ref{eq:Ham2}) represent the temperature($T$)-dependent singlet formation energy, and the singlet-singlet interaction strength, respectively. These appear as purely classical terms in the effective single-particle Hamiltonian. The second term describes the interaction of classical spins, ${\mathbf{\cal S}}_i$, with conduction electrons. The third term indicates that if a singlet exists at a given site, then another conduction electron can only appear in the triplet state which is higher in energy by $J_{\rm K}$.

\begin{figure}[t]
    \makebox[\columnwidth]{
        \includegraphics[width=0.98\columnwidth]{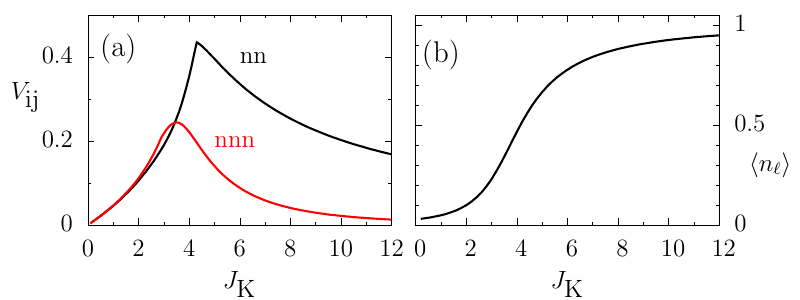}
    }
    \caption{(Color online)~ (a) Nearest-neighbor and next-nearest-neighbor interaction energies between Kondo singlets, and (b) density of electrons, $\langle n_{\ell} \rangle$ at the Kondo impurity site as a function of $J_{\rm K}$. The results are obtained via ED on $6 \times 6$ cluster.}
    \label{fig1}
\end{figure}

The interaction energy, $V_{ij}$, between singlets in the Hamiltonian Eq. (\ref{eq:Ham2}) is estimated via ED of the Hamiltonian Eq. (\ref{eq:Ham1}) in the limit of two-impurities and two-electrons \cite{Jones1987, Neel2011, Trishin2023, Spinelli2015}. Assuming that the impurities are located at sites $i$ and $j$, we find the exact ground state energy $E^{(2)}_{ED}(r_{ij})$ for a given $J_{\rm K}$, where the superscript refers to the number of impurities present. The interaction energy between singlets located at sites $i$ and $j$ is identified via $V_{ij} = E^{(2)}_{ED}(r_{ij}) - 2E^{(1)}_{ED}$, where $E^{(1)}_{ED}$ is the energy for the case of one impurity and one electron. The ED results are obtained on a $N = 6^2$ lattice with periodic boundary conditions (See Supplemental Material). The singlet-singlet interaction strength displays a non-monotonic behavior with $J_{\rm K}$ (see Fig. \ref{fig1} (a)). The definition of $V_{ij}$ is motivated by the observation that for sufficiently large separation $r > r_0$ between sites $i$ and $j$, $E^{(2)}_{ED}(r) \approx 2E^{(1)}_{ED}$. In fact, the value of $r_0$ provides an estimate for the radius of the Kondo cloud, which decreases upon increasing $J_{\rm K}$. This may also be inferred from the variation in the electronic density at the impurity site, $\langle n_{\ell} \rangle$, which displays a crossover behavior between weak-screening limit at small $J_{\rm K}$ and a site-localized Kondo singlet (see Fig. \ref{fig1} (b)). Note that $\lambda_i \equiv 0$ leads to a commonly studied limit of the model that assumes the spins as classical vectors \cite{Dagotto2002}.

{\it Hybrid Monte Carlo method:--}
The Hamiltonian Eq. (\ref{eq:Ham2}) describes lattice Fermions interacting with classical spins in the presence of impurity potentials. More importantly, variables $\{\lambda_i\}$ that determine the fractional contribution towards singlet energy at each site are to be dynamically determined along with a constraint $0 \leq \lambda_i \leq 1$. For a consistent description, $\lambda_i$ represents the electronic charge trapped in singlet formation around site $i$, and $N_{\rm K} = \sum_i \lambda_i$ is the the total number of Kondo singlets. Note that the proposed Hamiltonian contains terms that are either purely classical or bilinear in electronic operators. A reliable and unbiased approach to study this class of Hamiltonians is HMC. The problem of finding the ground state of the Hamiltonian reduces to that of optimizing the values of $\lambda_i$ and ${\mathbf {\cal S}}_i$. We perform this optimization via a Markov chain HMC where time-independent Schrodinger equation is solved at each step of Monte Carlo update (see Supplemental Material). 
The filling fraction of conduction band is a dynamical variable in the Monte Carlo updates, and is given by $n_{\rm c} = n^0_{\rm c} - n_{\rm K}$, where $n_{\rm K} = N_{\rm K}/N$ is the fraction of singlets in a given configuration and $n^0_{\rm c}$ is the filling fraction of the KLM. 

The RKKY driven magnetic order is characterized by $\lambda_i \equiv 0$. On the other hand, $\lambda_i \equiv 1$ represents a fully Kondo screened phase, which is the large $J_{\rm K}$ limit of the Doniach phase diagram. Our approach naturally allows for possibilities that are beyond the homogeneous phases. This is realized when $\lambda_i$ vanishes on a fraction of sites. Most importantly, the HMC approach determines the pattern of sites with finite $\lambda_i$ in an unbiased manner. We supplement the HMC simulations, which are limited to small clusters, with the variational calculations in order to minimize the finite-size effects. The variational ansatz is motivated by the HMC results, and the corresponding computations are carried out on much larger lattices by utilizing the super-cell structure of the variational configurations (see Supplemental material).

\begin{figure}[t]
    \makebox[\columnwidth]{
        \includegraphics[width=0.98\columnwidth]{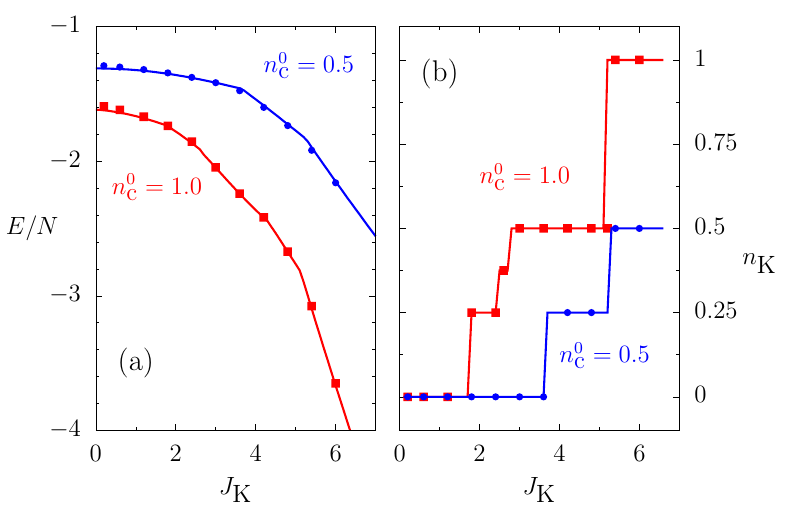}
    }
    \caption{(Color online)~ (a) Ground state energy per site, and (b) fraction of Kondo singlets, as a function of $J_{\rm K}$ obtained via the variational calculation (solid lines) and via the HMC simulations (symbols). The results for both, the half ($n^0_{\rm c} = 1$) and the quarter($n^0_{\rm c} = 0.5$) filled conduction bands are shown.}
    \label{fig2}
\end{figure}

{\it Partially magnetically ordered states:--}
We perform HMC simulations on a square lattice with $N=8^2$ sites and periodic boundary conditions. The ground state energy comparison between HMC simulations and the variational calculations for $n^0_{\rm c} = 1$ and $n^0_{\rm c} = 0.5$ is displayed in Fig. \ref{fig2} (a). The variational calculations of energy per site, obtained for a momentum space grid with $N=1024^2$ points in the first Brillouin zone, show a good agreement with the HMC results. The fraction of Kondo singlets, $n_{\rm K}$, already serves as an indicator of the partial order. The HMC simulations, as well as the variational calculations, display a plateau structure in the variation of $n_{\rm K}$ for both half- and quarter-filling (see Fig. \ref{fig2} (b)). Presence of plateau at certain fractions of Kondo singlets is indicative of the stability of the corresponding PMO phases. 
We provide a detailed characterization of the PMO phases via spin and singlet structure factors (SFs) defined respectively as,
\begin{eqnarray}
S({\bf q}) & = & \frac{1}{N^2}\sum_{ij} {\bf {\cal S}}_i \cdot {\bf {\cal S}}_j~e^{{\rm i} {\bf q}\cdot ({\bf r}_i - {\bf r}_j)},   \nonumber  \\ 
X({\bf q}) & = & \frac{1}{N^2}\sum_{ij} {\lambda}_i {\lambda}_j~e^{{\rm i} {\bf q}\cdot ({\bf r}_i - {\bf r}_j)}.
\label{eq:SF}
\end{eqnarray}

\begin{figure}[t]
    \makebox[\columnwidth]{
        \includegraphics[width=0.94\columnwidth]{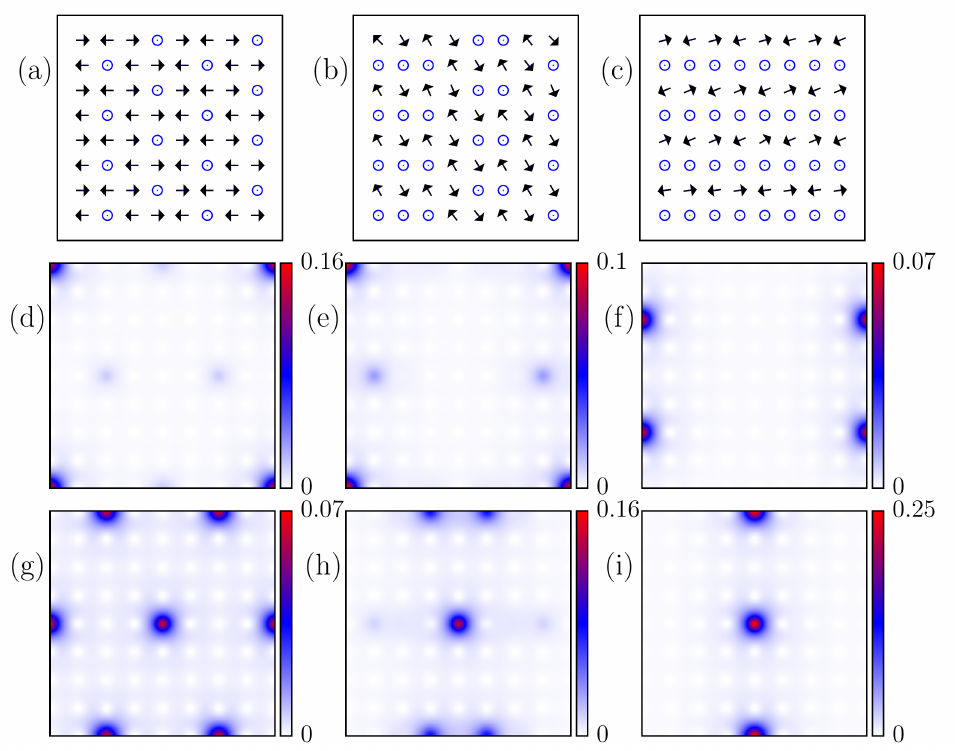}
    }
    \caption{(Color online)~ (a)-(c) Real space configurations of ${\cal S}_i$ (arrows) and $\lambda_i$ (circles) for, (a) $J_{\rm K} = 2.4$, (b) $J_{\rm K } = 2.6$, and (c) $J_{\rm K } = 4.2$ obtained at low temperature via HMC. The corresponding spin SFs (d)-(f), and singlet SFs (g)-(i). The results shown here are for half-filled KLM.}
    \label{fig3}
\end{figure}

We begin by describing the evolution of ground states at half-filling upon increasing the strength of the Kondo coupling. The ground state for $J_{\rm K}/t < 1.8 $ is Neel antiferromagnet. This is followed by three distinct PMO ground states that exist in the range $1.8 < J_{\rm K} < 5.3 $. The real-space configurations for these three phases are shown in Fig. \ref{fig3} (a)-(c). A common feature of these states is the presence of a connected hopping pathway of the sites with magnetic moments. This allows the system to gain kinetic energy via the conduction electrons, as well as the correlation energy from the Kondo singlets. 

The spin SF peaks at ${\bf q} = (\pi,\pi)$ in the Neel ground state as expected (see Supplemental Material). The magnetic sub-lattice for the PMO states is best visualized as a magnetic lattice with ordered vacancies. This can either lead to secondary peaks in the spin SF for the PMO states (see Fig. \ref{fig3} (d)-(e)) or to a complete shift of the primary peak locations (see Fig. \ref{fig3} (f)). 
The magnetic order obtained in these PMO states cannot be captured within the regular single- or multi-Q ansatz that is commonly employed to describe magnetic ordering in the semiclassical treatment of the KLM. The arrangement of singlets in the PMO states can also be identified via the singlet SF (see Fig. \ref{fig3} (g)-(i)). Note that the peak at ${\bf q} = 0$ in the singlet SF carries the information about the fraction of singlets present in the state, which can be read from Fig. \ref{fig2} (b).

\begin{figure}[t]
    \makebox[\columnwidth]{
        \includegraphics[width=0.98\columnwidth]{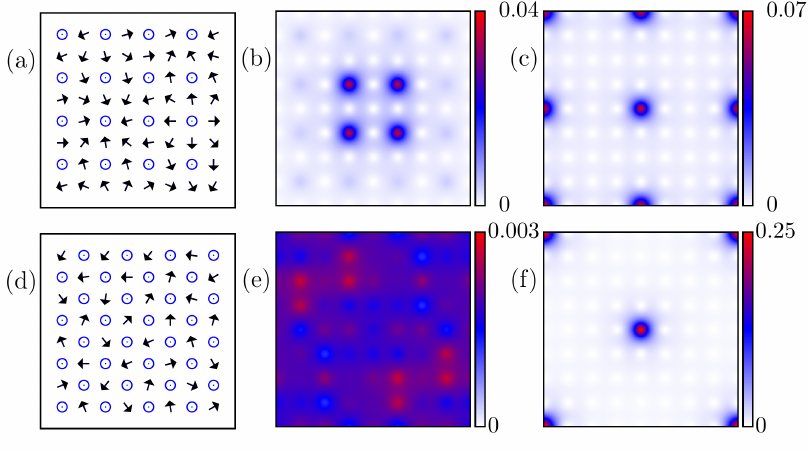}
    }
    \caption{(a) Real space configuration, (b) spin SF and (c) singlet SF at low temperature obtained via HMC for $J_{\rm K} = 4.2$. (d)-(f) Same quantities as (a)-(c), respectively, for $J_{\rm K} = 6.0$. The results shown here are for quarter-filled KLM.} 
    \label{fig4}
\end{figure}

At quarter filling, the ground state for small $J_{\rm K}$ supports pure magnetic order with spiral wave-vectors at $(\pm \pi, 0)$ and $(0, \pm \pi)$ (see Supplemental). With increasing $J_{\rm K}$, the ground state becomes PMO with the magnetic sub-lattice forming a Lieb lattice (see Fig. \ref{fig4} (a)) and the corresponding spin SF displaying non-trivial peaks at ${\bf q} = (\pm \pi/4, \pm \pi/4)$ (see Fig. \ref{fig4} (b)). The singlet order in this state is depicted via the peaks at $(0,\pi)$,$(\pi,0)$ and $(\pi,\pi)$ in the singlet SF (see Fig. \ref{fig4} (c)). At large $J_{\rm K}$, a staggered arrangement of magnetic and singlet sites describes the ground state (see Fig. \ref{fig4} (d), (f)). In this state all the charge carriers are bound in the Kondo singlets and therefore the RKKY mechanism to induce order in the magnetic sublattice is inactive. Hence, this Kondo insulating state supports a peculiar mixed classical- and quantum paramagnetic behavior. The lack of magnetic order in this state is apparent from the absence of sharp peaks in the spin SF (see Fig. \ref{fig4} (e)).

\begin{figure}[t]
    \makebox[\columnwidth]{
        \includegraphics[width=0.98\columnwidth]{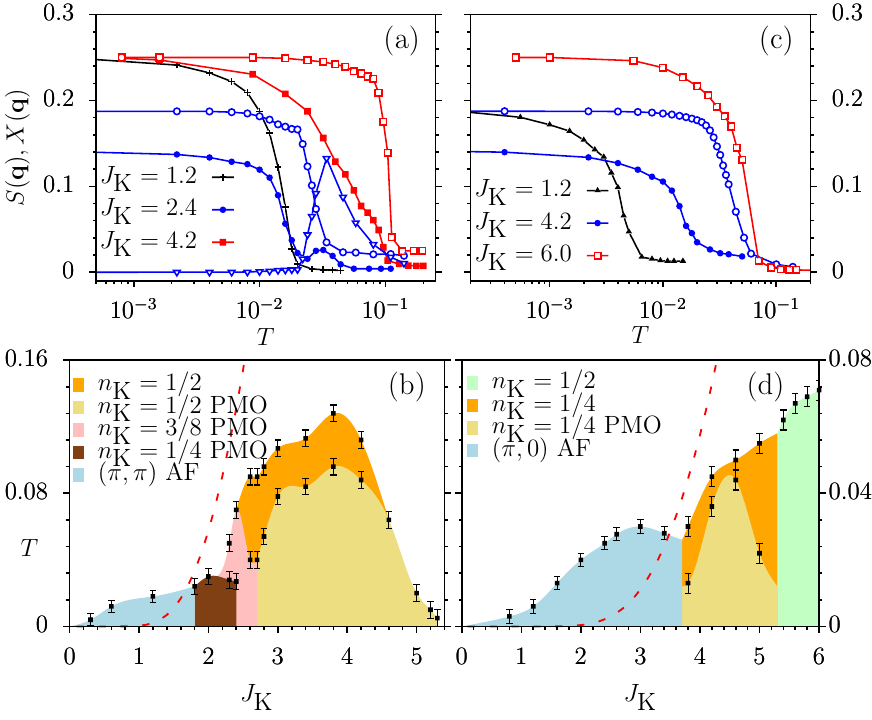}
    }
    \caption{(Color online)~ Temperature dependence of components of spin and singlet SFs for representative values of $J_{\rm K}$ at, (a) half-filling and (c) quarter filling. The closed (open) symbols correspond to the spin (singlet) SFs. Panel (a): S(($\pi,\pi$)) at $J_{\rm K} = 1.2$; $S((\pi,\pi))$, $X((\pi/2,\pi))+X((\pi,0))$ (circles) and $X((\pi/4,\pi ))$ (triangles) at $J_{\rm K} = 2.4$; $S((\pi,\pi/2))$ and $X((0, \pi))$ at $J_{\rm K} = 4.2$. Panel (c): $S((\pi,0))$ at $J_{\rm K} = 1.2$; $S((\pi/4, \pi/4))$ and $X((\pi, 0)) + X((0, \pi)) + X((\pi, \pi))$ at $J_{\rm K} = 4.2$; $X((\pi, \pi))$ at $J_{\rm K} = 6.0$. $T$-$J_{\rm K}$ phase diagrams for, (b) half-filled and (d) quarter-filled bands. The red dashed lines in (b) and (d) are the exponential Kondo temperature scales.}
    \label{fig5}
\end{figure}

An important advantage of the HMC simulations is the direct access to the finite-temperature properties.
The variations of relevant components of spin and singlet SFs help us organize finite temperature phase diagrams. Given that the results are obtained on finite clusters, we estimate the ordering temperature from the inflexion point in the temperature dependence of the relevant components of the SFs. For the case of half-filling, the ground state is Neel antiferromagnet for $J_{\rm K} < 1.8$, and an $n_{\rm K} = 1/4$ PMO state shown in Fig. \ref{fig3}(a) for $1.8 < J_{\rm K} < 2.4$. For $2.1 < J_{\rm K} < 2.4$, the transition from the high-$T$ paramagnetic to the $n_{\rm K} = 1/4$ ground state proceeds via an intermediate $n_{\rm K} = 3/8$ phase. This is inferred from the presence of peaks in the singlet SF at ${\bf q} = (\pm \pi/4,\pm \pi)$ in the intermediate $T$ regime, and at ${\bf q} = (\pm \pi/2,\pm \pi)$ and $(\pm \pi,0)$ at low $T$ (see Fig. \ref{fig5}(a),(b)). The $n_{\rm K} = 3/8$ phase, shown in Fig. \ref{fig3} (b), is the ground state for $2.4 < J_{\rm K} < 2.7$. For $J_{\rm K}>2.7$, a stripe-like magnetic and singlet order with $n_{\rm K} = 1/2$ becomes the ground state. Interestingly, an intermediate phase that lacks magnetic order but retains singlet order separates the ground state from the high-$T$ paramagnetic state. We summarize the above results as a $T-J_{\rm K}$ phase diagram at half-filling (see Fig. \ref{fig5} (b)). 

A similar evolution with $J_{\rm K}$ of the magnetic phases is obtained at quarter filling. The ground state changes from purely magnetic to the Kondo-correlated non-magnet via an intermediate $n_{\rm K}=1/4$ PMO phase existing for $3.7 < J_{\rm K} < 5.3 $, and characterized in Fig. \ref{fig4}(a)-(c). Note that in this state the singlets reside on a square lattice with doubled lattice spacing. Consequently, the magnetic sites define a Lieb lattice with emergent flat bands for itinerant fermions. This emphasizes a general principle that, while a mass-enhancement is a useful way to describe transport in the KLM, a complete band reconstruction is essential to describe some of the ordered phases. The separation of onset-temperature scales for singlet and spin SFs (see Fig. \ref{fig5} (c)) for $J_{\rm K} = 4.2$ indicates the presence of a novel paramagnetic phase at intermediate temperatures where the singlets remain ordered while the Lieb sublattice loses its magnetic order. For $J_{\rm K} > 5.3$, a non-magnetic ground state with singlet order is obtained as the ground state (see Fig. \ref{fig4}(d)-(f)). In this phase, the singlets reside on a $45^{\circ}$-tilted square lattice and the magnetic sublattice remains paramagnetic. An identical checkerboard ordering of Kondo-screened sites at quarter filling has also been reported in a cDMFT study \cite{Peters2017}. While this simple picture of coexisting singlets and paramagnetic moments is likely to be unstable at low $T$ due to additional quantum effects, a symmetry-restored version is a possible candidate for the ground state. We find that all the PMO states obtained in our study have a common feature: they consist of a connected network of magnetic sites with a corresponding arrangement of Kondo singlets (see Fig. \ref{fig3} (a)-(c) and Fig.\ref {fig4} (a)). We note that many of the experimentally reported PMO states also consist of a connected network of  magnetic sites \cite{Iyer2023, Lucas2017, Movshovich1999}. This has previously been interpreted as a consequence of geometrical frustration. We predict that such PMO phases also exist in Kondo lattices without geometrical frustration as the mechanism for their stability has origin in the kinetic energy of conduction electrons. 

{\it Conclusion:--}
Using a semiclassical effective model Hamiltonian that combines crucial features of the KLM in the weak and strong coupling limits, we have shown that PMO phases exist as the ground states in the intermediate coupling regime. We find three distinct PMO states with $n_{\rm K} = 1/4, 3/8, 1/2$ at half filling and a state with $n_{\rm K} = 1/4$ at quarter filling. In addition, we identify states with singlet order and no magnetic order in the intermediate to large $J_{\rm K}$ regime. Our approach is qualitatively different from the existing mean-field approximations in that it emphasizes on treating the complex spatial correlations present in the model. Indeed, a common feature of all the PMO states is to have the Kondo singlets organized in a pattern that allows for a connected network of magnetic sites. This suggests that the effective single-particle electronic dispersion in a Kondo system can also be modified in manner that is not described by a simple mass enhancement. This is particularly likely for the KLM near half filling where the bare non-interacting band cannot be approximated by a quadratic dispersion. Therefore, we have identified a band-reconstruction mechanism for partial order in Kondo systems which is inherently non-local in nature. Note that this is completely unrelated to a local mechanism of partial ordering that originates from, for example, geometrical frustrations. Some of the PMO states obtained in our work share common features with those reported in experiments \cite{Lucas2017, Iyer2023, Movshovich1999}, and in previous theoretical investigations \cite{Peters2017, Costa2017, Motome2010a}. Our results suggest a two-fold extension of the well-known Doniach phase diagram: (i) a regime of PMO ground states for intermediate strength of Kondo coupling, (ii) an intermediate-$T$ regime of quantum-correlated non-magnetic phases for large $J_{\rm K}$. The concept of ssSSB introduced in this work may also be applicable to other models of correlated Fermionic or Bosonic systems. 
Suitable generalizations of the effective Hamiltonian approach presented in this work can be particularly useful for investigating the intermediate coupling regime of models of interacting quantum particles.

{\it Acknowledgments:--}
We acknowledge the use of High Performance Computing facility at IISER Mohali.

\bibliographystyle{apsrev4-1} 
\bibliography{ssSSB_v1}

\begin{widetext}


\section*{Supplemental Material}
\setcounter{section}{0}
\setcounter{equation}{0}
\setcounter{figure}{0}
\renewcommand\thesection{\Alph{section}}

\renewcommand{\theequation}{S\arabic{equation}}
\renewcommand{\thefigure}{S\arabic{figure}}

\section{Exact Diagonalization and inter-singlet interactions}
 
Consider two spin-1/2 impurities Kondo-coupled to tight-binding  electrons on a square lattice, specified by the Hamiltonian, 
\begin{eqnarray}
	H & = & - t \sum_{\langle ij \rangle,\alpha} (c^\dagger_{i\alpha} c^{}_{j\alpha} + {\textrm H.c.})
	+ J_{\text{K}} \sum_{i \in \{\ell ,m\}} {\bf S}_i \cdot {\bf s}_i, 
	\label{eq:S1}
\end{eqnarray}
\noindent
also given in Eq. 1. in the main text. The various symbols have the same meaning as assigned in the main text. In addition to the electronic kinetic energy, the Hamiltonian consists of the Kondo coupling between the spins of conduction electrons and those localized at sites $\ell$ and $m$. We consider the case of two conduction electrons and represent the Hamiltonian in the full Hilbert space. The states of the Hilbert space can be represented as a product of states of the conduction electrons and those of localized spins. It is useful to organize the states of the conduction electrons in terms of the spin orientations of the two particles. This leads to  $N(N-1)/2$ states having the spins of both the conduction electrons $\uparrow$, where $N= n_x \times n_y$ is the total number of lattice sites. Similarly, there are $N(N-1)/2$ states having spins of both conduction electrons $\downarrow$. Finally, there are $N^2$ states of the conduction electrons that have one particle with spin $\uparrow$ and other with spin $\downarrow$. Thus, the total number of states of two conduction electrons on $N$ sites is $N(2N-1)$. Including the $4$ states of two spin-$1/2$ impurities at sites $\ell$ and $m$$, $$\{\uparrow \uparrow, \uparrow\downarrow, \downarrow\uparrow \text{and} \downarrow\downarrow\}_{\ell,m}$, the size of the full Hilbert space is $4 \times N(2N-1)$. We represent the Hamiltonian as a matrix in this Hilbert space and diagonalize it using LAPACK libraries. For these computations, we considered a $6 \times 6$ square lattice with periodic boundary conditions. For a given $J_{\rm K}$, we calculate the ground state energy($E^{(2)}_{ED}(r_{\ell m})$) for different separations, $r_{\ell m}$, of the impurity spins. In order to estimate the inter-singlet interaction energy, we require the energy of the state with infinite separation between the singlets. A good estimate for this energy is to consider two isolated systems, each with one conduction electron and one localized impurity spin. Therefore, we take $E^{(2)}_{ED}(r_{\ell m} \rightarrow \infty) \approx 2E^{(1)}_{ED}$. Finally, we estimate the inter-singlet interaction energy as, $V_{\ell m} = E^{(2)}_{ED}(r_{\ell m}) - 2E^{(1)}_{ED}$. 
The above procedure is repeated for different $J_{\rm K}$ values. This allows us to plot the interaction energy as a function of  $J_{\rm K}$, as shown in Fig. 1(a) in the main text.

\section{Hybrid Monte Carlo Simulations}

The effective Hamiltonian (Eq. 2 in the main text)  put forward in this work consists of classical as well as quantum variables. The classical degrees of freedom are: (i) localized magnetic moments, ${\cal S}_i $ and (ii) probability factors, $\lambda_i$. The quantum variables are denoted by the creation and annihilation operators $c^{}$, $c^{\dagger}$ for the conduction electrons.
Within the semiclassical approximation already employed in considering the localized spins as classical, the Hamiltonian is bilinear in the electronic operators. Therefore, the many-particle problem simplifies drastically as we can build the multi-particle states from the single-particle states. Furthermore, the Hilbert space for single-particle states grows linearly with the system size.  However, the ground state of the system can only be obtained if we have the knowledge of the classical configuration that would minimize the total energy. The most suited method to tackle this class of Hamiltonians is the Hybrid Monte Carlo (HMC).
The methods proceeds similar to the classical Monte Carlo approach for a classical Hamiltonian, with a crucial difference that the energy of the classical configuration also depends of the fermionic free energy in that configuration. Therefore, a solution of a single particle Schrodinger Equation is required at every Monte Carlo update step. Since the diagonalization time scales as $N^3$ with the number of sites, and the Monte-Carlo sweep of the entire system requires $N$ steps, the HMC simulations scale as $N^4$.

\begin{figure}[h]
\includegraphics[width=.8 \columnwidth,angle=0,clip=true]{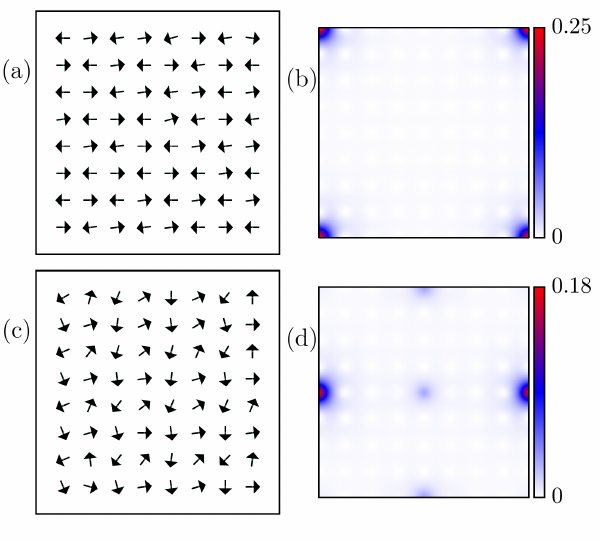}
\caption{(a) Real space configuration, and (b) spin SF at low-$T$ obtained via HMC for $J_{\rm K} = 1.2$ at half filling. The conventional Neel antiferromagnetic state is the ground state for small $J_{\rm K}$.
(c) Real space configuration, and (d) spin SF at low-$T$ obtained via HMC for $J_{\rm K} = 1.2$ at quarter filling.
}
\label{figS1}
\end{figure}

In the present study, we consider $8\times 8$ square lattice with periodic boundary conditions. The diagonalization of the fermionic problem is carried out using the CHEEVX subroutine of the LAPACK library. We have taken $10^4$ Monte Carlo steps for the equilibriation and a similar number of steps for calculating the observables. We start our simulation at high temperature with a random initial state for both ${\cal S}_i$ and $\lambda_i$. We anneal the system by lowering the temperature in a sufficiently large number of steps.

\section{Variational Calculations}

\begin{figure}[h]
\includegraphics[width=.7 \columnwidth,angle=0,clip=true]{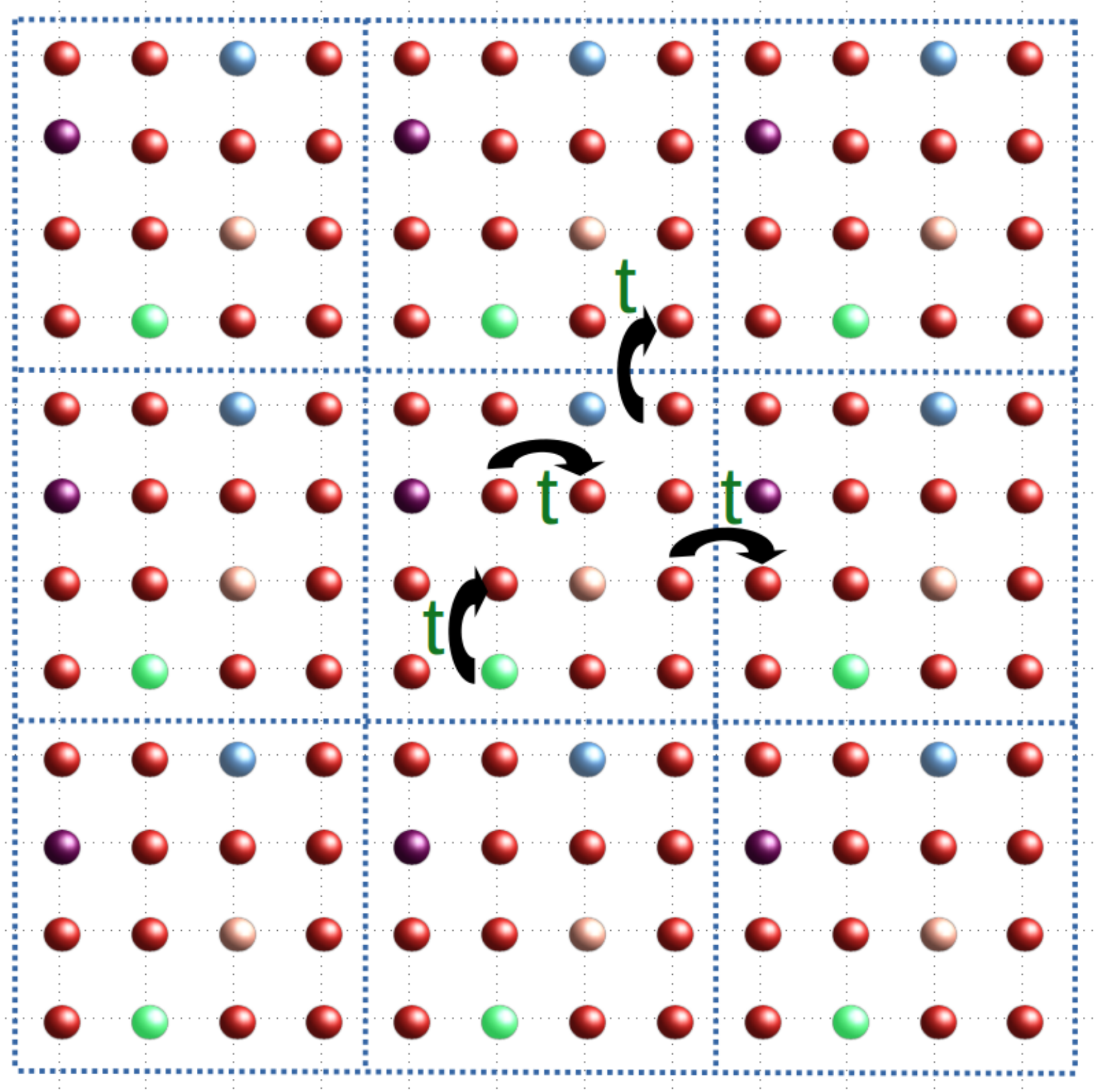}
\caption{ Schematic view of a square lattice with super-cell structure. Note that the color pattern of sites, which is representative of on-site potential values, is repeated across cells. On Fourier transforming the hopping Hamiltonian, the inter-cell hopping terms would appear with twisted boundary conditions (see text).  
}
\label{figS2}
\end{figure}

In order to minimize the finite-size effects for the identification of the ground states, we set up variational calculations on larger lattices. The choice of candidate states of the variational calculations is inspired by the results of the HMC simulations. We consider a square lattice of size $n_x \times n_y$, and a general block-periodic configuration with block size  $c_x \times c_y$. For illustration purpose, we consider a spinless tight-binding Hamiltonian on a square lattice with additional on-site potential term $\sum_i \epsilon_i c^{\dagger}_ic^{}_i$. The pattern of $\epsilon_i$ follow a block-periodic structure. A schematic configuration with $n_x = n_y = 12$ and $c_x = c_y =4$ is shown in Fig. \ref{figS2}. It is easy to show that if we Fourier transform the single-particle Hamiltonian for such block-periodic configurations then in $k_x, k_y$ space we are required to diagonalize a matrix of dimension $c_x\times c_y$ at each $k_x, k_y$ point. While the local terms in the Hamiltonian remain invariant under the Fourier transformation, and hence independent of $k_x, k_y$, the hopping terms acquire a twisted boundary condition. For the variational calculation performed in this work, we use $n_x = n_y = 1024$ and various choices of  $c_x \times c_y$. The resulting Hamiltonian matrix for a given $k_x, k_y$ point is given by,  
\begin{eqnarray}
	H & = & -\sum_{\langle k_i, k_j \rangle} t(k_x, k_y)(c^\dagger_{k_i} c^{}_{k_j} + {\textrm H.c.})
	+ \sum_{k_i} \epsilon_i ~ c^{\dagger}_{k_i}c^{}_{k_i}, 
	\label{eq:S1}
\end{eqnarray}
\noindent
where the summation is over the nearest neighbor sites $k_i, k_j$ of the $c_x \times c_y$ grid with inequivalent sites. The hopping parameters are such that $t(k_x, k_y) = t~e^{ik_{x(y)}}$, for hopping across the boundaries on $x(y)$ edges, and $t(k_x, k_y) = t$, otherwise. The use of hoppings with $k_x, k_y$ dependent phase factors is also sometimes referred to as twisted boundary conditions.

\begin{figure}[h]
    \makebox[\columnwidth]{
        \includegraphics[width=0.48\columnwidth]{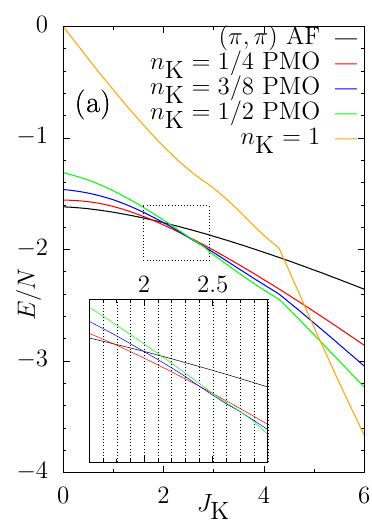}
        \includegraphics[width=0.48\columnwidth]{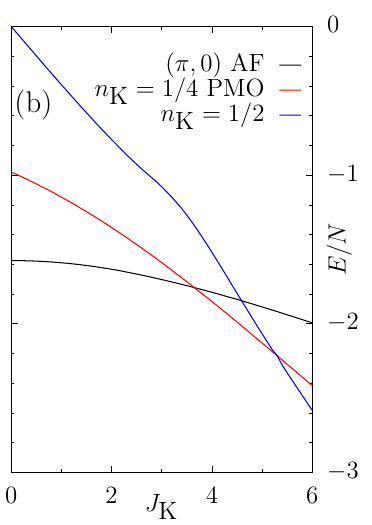}
    }
    \caption{Energy comparison of different variational states for, (a) half and (b) quarter filling. Inset in (a) shows the zoomed-in view of the dotted box.}
    
\label{figS3}
\end{figure}

The idea illustrated above for a spinless tight-binding model can be generalized in a straightforward manner to the effective single-particle Hamiltonian used in this work. The single particle spectrum can be obtained by diagonalizing $(n_x n_y)/(c_x c_y)$ matrices of dimension $2c_xc_y$ each, instead of a single matrix of dimension $2n_xn_y$. In Fig. \ref{figS3}(a)-(b) we show the energies of the variational phases as a function of Kondo coupling for the half- and quarter-filled case. The transition points between the magnetic and different PMO states are identified by the crossing of the energy curves.

\section{Density of States}
As discussed in the main text, the PMO states are characterized by specific ordered arrangements of singlets and magnetic sites on the lattice. Interestingly, the stabilization of the PMO states is itself dictated by the effective single-particle band structure. This band-reconstruction mechanism for partial order is one of the key findings of the present work. Here, we show some examples of how the effective single-particle band structure gets drastically modified in the PMO states. Having obtained the energy spectra using variational approach, we calculate the density of states(DOS) using,
\begin{eqnarray}
{\cal N}(\omega) = \dfrac{\gamma}{\pi} \sum_{\mathbf{k}} \dfrac{1}{\gamma^2 + (\omega - \epsilon(\mathbf{k}))^2},
\end{eqnarray}
\noindent
where $\gamma$ is the Lorentzian broadening. In our calculation, we have used $\gamma = 0.01$. In Fig. \ref{figS4} and Fig. \ref{figS5}, we show the DOS for different magnetic and PMO states at half- and quarter-filling.

\begin{figure}[h]
\includegraphics[width=.9 \columnwidth,angle=0,clip=true]{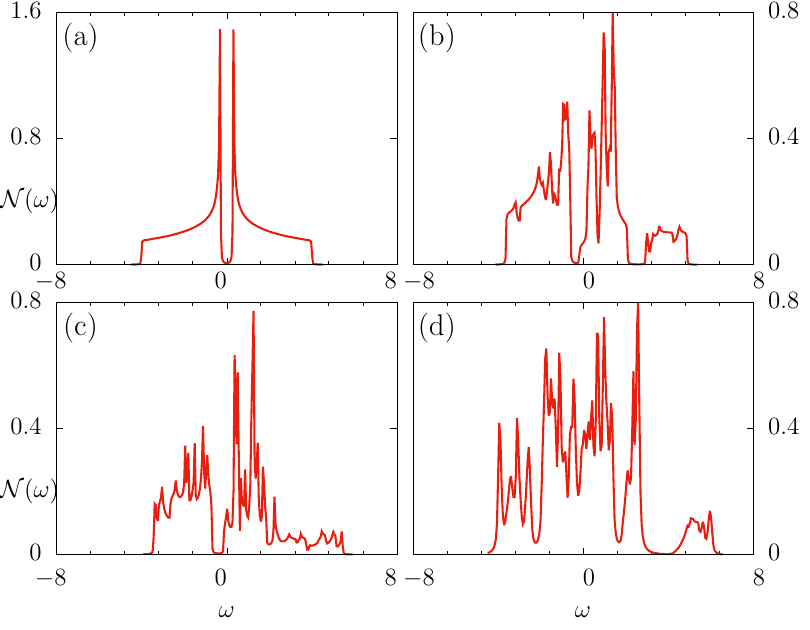}
\caption{Density of states at half-filling for (a) $J_{\rm K} = 1.2$, (b) $J_{\rm K } = 2.4$, (c) $J_{\rm K } = 2.6$, and (d) $J_{\rm K } = 4.2$. Note that the underlying partial order is different in each case, and it leads to non-trivial effects in the DOS. 
}
\label{figS4}
\end{figure}

\begin{figure}[h]
\includegraphics[width=.98 \columnwidth,angle=0,clip=true]{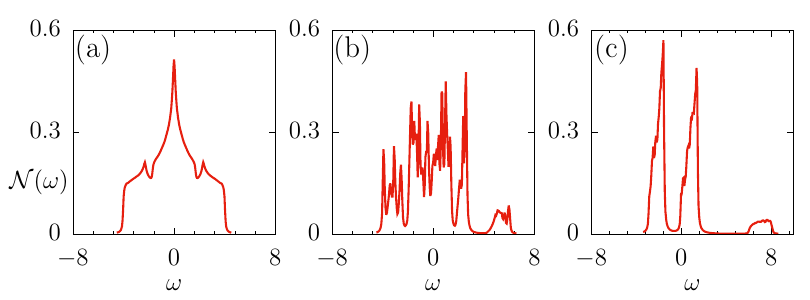}
\caption{Density of states at quarter-filling for, (a) $J_{\rm K} = 1.2$, (b) $J_{\rm K } = 4.2$, and (c) $J_{\rm K } = 6.0$. 
}
\label{figS5}
\end{figure}
\end{widetext}

\end{document}